\documentclass[pre,floats,twocolumn,showpacs,floatfix,superscriptaddress]{revtex4}
\usepackage{graphics,graphicx,dcolumn,bm,fleqn,epic,eepic,float}
\usepackage{amssymb,amsmath,multirow,rotate,color,float}
\usepackage[latin1]{inputenc}

\usepackage[dvips]{epsfig}

\def\p{\par $\bullet$ }

\begin{document}

\title{Drag force on a sphere moving towards
an anisotropic super-hydrophobic plane}

\author{Evgeny S. Asmolov}
\affiliation{A.N.~Frumkin Institute of Physical
Chemistry and Electrochemistry, Russian Academy of Sciences, 31
Leninsky Prospect, 119991 Moscow, Russia}
\affiliation{Institute of Mechanics, M. V. Lomonosov Moscow State University, 119992 Moscow, Russia}
\affiliation{Central Aero-Hydrodynamics Institute, 1 Zhukovsky str., Zhukovsky, Moscow region, 140180, Russia}

\author{Aleksey V. Belyaev}
\affiliation{A.N.~Frumkin Institute of Physical
Chemistry and Electrochemistry, Russian Academy of Sciences, 31
Leninsky Prospect, 119991 Moscow, Russia}
\affiliation{Department of Physics, M. V. Lomonosov Moscow State University, 119991 Moscow, Russia}

\author{Olga I. Vinogradova}
\affiliation{A.N.~Frumkin Institute of Physical
Chemistry and Electrochemistry, Russian Academy of Sciences, 31
Leninsky Prospect, 119991 Moscow, Russia}
\affiliation{Department of Physics, M. V. Lomonosov Moscow State University, 119991 Moscow, Russia}
\affiliation{DWI, RWTH Aachen, Pauwelsstr. 8,
52056 Aachen, Germany}

\newcommand\Xsin{\mbox{sin}}
\newcommand\Xcos{\mbox{cos}}
\newcommand\Xsec{\mbox{sec}}
\newcommand\Xlog{\mbox{ln}}

\newcommand*{\mycommand}[1]{\texttt{\emph{#1}}}

\begin{abstract}

We analyze theoretically a high-speed drainage of liquid films squeezed
between a hydrophilic sphere and a textured super-hydrophobic plane, that contains trapped gas bubbles. A super-hydrophobic wall is characterized by parameters $L$ (texture characteristic length), $b_{1}$ and $b_2$ (local slip lengths at solid and gas areas), and $\phi_{1}$ and $\phi_2$ (fractions of solid and gas areas). Hydrodynamic properties of the plane are fully expressed in terms of the effective slip-length tensor  with eigenvalues that depend on texture parameters and $H$ (local separation). The effect of effective slip is predicted to decrease the force as compared with expected for two hydrophilic surfaces and described by the Taylor equation. The presence of additional length scales, $L$, $b_1$ and $b_{2}$,  implies that a film drainage can be much richer than in case of a sphere moving towards a hydrophilic plane.  For a large (compared to $L$) gap the reduction of the force is small, and for all textures the force is similar to expected when a sphere is moving towards a smooth hydrophilic plane that is shifted down from the super-hydrophobic wall. The value of this shift is equal to the average of the eigenvalues of the slip-length tensor. By analyzing striped super-hydrophobic surfaces, we then compute the correction to the Taylor equation for an arbitrary gap. We show that at thinner gap the force reduction becomes more pronounced, and that it depends strongly on the fraction of the gas area and local slip lengths. For small separations we derive an exact equation, which relates a correction for effective slip to texture parameters.  Our analysis provides a framework for interpreting recent force measurements in the presence of super-hydrophobic surface.

\end{abstract}


\maketitle

\section{Introduction}

Super-hydrophobic Cassie (SH) surfaces have opened a whole new field of investigation,
with both fundamental and practical perspectives~\cite{quere.d:2005}. They are able to trap air at the liquid-solid
interface, leading to remarkable (`super') properties, such as a very large water contact angle and low hysteresis. This strong hydrophobicity has macroscopic implications in the context of self-cleaning~\cite{blossey.r:2003}
 and impact processes~\cite{richard.d:2000,tsai.p:2010}.  SH surfaces
 could also revolutionize microfluidic lab-on-a-chip systems~\cite{stone2004,squires2005} since the large effective slip of SH surfaces~\cite{ybert.c:2007,feuillebois.f:2009,vinogradova.oi:2011} compared to smooth hydrophobic channels~\cite{vinogradova.oi:2009,vinogradova.oi:2003,charlaix.e:2005} can greatly lower the viscous drag. SH surfaces can also amplify electrokinetic pumping~\cite{Squires08,bahga:2009,belyaev.av:2011a} and mixing~\cite{feuillebois.f:2010b,vinogradova.oi:2011} in microfluidic devices.

 \begin{figure}
 \includegraphics[width=6 cm]{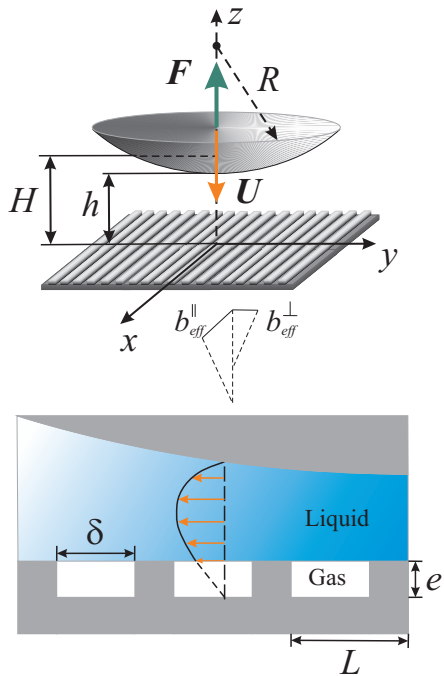}\\
  \caption{(Color online) Sketch of a hydrophilic sphere approaching a striped super-hydrophobic surface (top), and  its model representation as a flat interface with patterns of boundary conditions  (bottom).}
  \label{fig:geometry}
\end{figure}

 In addition, this superlubrication potential should dramatically modify a squeeze film drainage between surfaces.   In the previous study~\cite{belyaev.av:2010b} we have analyzed a drag force, $F$, on a hydrophilic disk approaching a SH wall (the Reynolds problem). Here we explore what happens when a hydrophilic sphere of radius $R$ is driven towards a SH plane with a velocity $U$ (see Fig.~\ref{fig:geometry}), i.e. we address the so-called Taylor problem. Beside its significance as a geometry of surface forces apparatus (SFA) and atomic force microscope (AFM) force experiments, it represents a typical situation of phenomena of `viscous adhesion', coagulation, and more. In case of hydrophilic surfaces a hydrodynamic force reads~\cite{note1}
 \begin{equation}\label{taylor}
     F_T=\frac{6\pi\mu U R^2}{h}
 \end{equation}
 when the gap $h\ll R$. Here $\mu$ denotes a fluid dynamic viscosity. In case of a hydrophobic plane, characterized by a constant slip length $b$ (the distance between the solid at which the flow profile extrapolate to zero) the Taylor expression has to be corrected for slip~\cite{vinogradova.oi:1995d,vinogradova.oi:1995a}
 \begin{multline}\label{fb0}
    f^{\ast}=\frac{F}{F_T}= \frac{1}{4} \left(1+\frac{3 h}{2 b} \left[ \left(1 + \frac{h}{4 b}\right)\times \right. \right. {} \\
   {} \left. \left. \times\ln\left(1+\frac{4 b}{h}\right)-1 \right] \right)
\end{multline}
The
factor $f^{\ast}$ associated with hydrodynamic slip can significantly decrease the hydrodynamic resistance force provided $h$ is of the order of $4b$ or smaller. Eq.(\ref{fb0}) is often used to infer the value of a slip length from the force experiment. Indeed, the dynamic SFA/AFM force measurements~\cite{chan.dyc:1985} are extremely accurate at the nanoscale as compared to direct flow profiling, or velocimetry~\cite{note2}. Therefore, it is possible to measure even nanometric slip lengths~\cite{vinogradova.oi:2003,charlaix.e:2005,honig.cdf:2007,wang.y:2010}. The advantage of this (hydrophobic vs. hydrophilic) geometry of configuration is that it allows to avoid the gas bridging and long-range attractive capillary forces~\cite{andrienko.d:2004}, which appear when we deal with interactions of two hydrophobic solids~\cite{yakubov:00,tyrrell.jwg:2001,parker.jl:1994}.

For these reasons it is attractive to consider the hydrodynamic interaction of a hydrophilic sphere with a SH surface. However, despite of its importance for force experiments and numerous applications, the quantitative understanding of the problem is still challenging. The heterogeneous nature of the SH texture makes difficult a precise discussion of the liquid flow past composite regions, especially when surface patterns are anisotropic. It would seem therefore appropriate to bring a more modern theoretical technique to bear on this problem. In this paper we present some results of a study of a force, acting on a sphere approaching to the SH wall. Our theory is based on the effective slip approach introduced originally to describe flows past a single interface or in a flat channel~\cite{stone2004,Bazant08,kamrin.k:2010,vinogradova.oi:2011}. In this approach, the effective slip is evaluated by averaging of a flow over the length scale of the experimental configuration being applied at the hypotetical smooth surface. Such a boundary condition mimics the actual one along the true heterogeneous SH surface, where gas pockets are stabilized with a rough wall texture. A corollary of this is that the effective hydrodynamic slip, ${\bf b}_{\rm eff}$, generally depends on the flow direction being a tensor~\cite{Bazant08} and that it also depends on the separation between surfaces~\cite{belyaev.av:2010b,vinogradova.oi:2011}. The concept of an effective tensorial slip has provided a great deal of insides into various factors that determines flow in a flat channel, and allowed one to obtain  very simple solutions of complex problems~\cite{Bazant08,vinogradova.oi:2011}. We shell see that it can be successfully applied to solve the Taylor problem in the presence of the SH surface. Our theory has the merit of yielding useful (approximate) analytical results as well as being very well suited to numerical work.

Our paper is arranged as follows: In Sec.II some general consideration concerning a description of a drainage of a liquid film confined between a hydrophilic sphere and arbitrary SH textured plane are presented. Here we also describe some universal asymptotic solutions,  valid for any texture. Sec.III contains analytical and numerical  results for striped SH surface. We conclude in Sec.IV with a discussion of our results and their possible relevance for force experiments. Appendix A contains a derivation of equations in a thin gap limit.

\section{General theory}

In this section we describe the theory of a hydrodynamic interaction of a sphere with an idealized SH surface in the Cassie state (sketched in Fig.~\ref{fig:geometry}),
where a liquid slab lies on top of the surface
roughness. The liquid/gas interface is assumed to be flat with no meniscus
curvature, so that the modeled super-hydrophobic surface
appears as a perfectly smooth with a pattern of
boundary conditions. The latter are taken as low partial slip ($b_1$) over solid/liquid
areas and as large partial slip ($b_2$) over gas/liquid regions. We denote
as $\delta$ a typical length
scale of gas/liquid areas. The fraction of solid/liquid
areas will be denoted $\phi_1=(L-\delta)/L$, and of gas/liquid area $\phi_2=1-\phi_1=\delta/L$.
In this idealization, some assumptions may have a possible influence
on the friction properties and, therefore, a hydrodynamic force. First, by assuming flat
interface, we have neglected an additional mechanism for a dissipation connected with the meniscus curvature~\cite{harting.j:2008,lauga2009,sbragaglia.m:2007}. Second, we ignore a possible transition towards impaled (Wenzel) state that can be provoked by additional pressure in the liquid phase~\cite{pirat.c:2008,reyssat.m:2008}.

The flow of liquid in the gap satisfies Stokes equations
\begin{equation}
\mu \frac{\partial^2 {\textbf{v}_\tau}}{\partial z^2} \simeq \nabla_\tau p,
\label{NS}
 \end{equation}
 \begin{equation} \frac {\partial v_z}{\partial z} + \nabla_\tau\cdot \textbf{v}_\tau = 0,
\label{conteq}
\end{equation}
 where
$\textbf{v}_{\tau}=v_x \textbf{e}_x + v_y \textbf{e}_y$ is the lateral velocity, $v_z$ is the normal velocity (directed towards the sphere), $p(x,y)$ is the local pressure, and
$\nabla_{\tau}$ is the differential operator in a plane $(x,y)$, given by
\begin{equation}\label{NablaTau}
  \nabla_{\tau}= \frac{\partial}{\partial x}\textbf{e}_x + \frac{\partial}{\partial y}\textbf{e}_y,
\end{equation}

The SH plane (generally anisotropic) exhibits uniform tensorial slip~\cite{Bazant08}, so that the boundary conditions are
\begin{equation}\label{BC_0}
  z=0: \quad \left(v_{\tau}\right)_i=\left(b_{\rm eff}\right)_{ij}\frac{\partial \left(v_{\tau}\right)_j}{\partial z}, \quad v_z=0;
\end{equation}
\begin{equation}\label{BC_H}
  z=H(x,y): \quad \textbf{v}=-U \textbf{e}_z,
\end{equation}
with summation over repeated indices. Near the axis the surface of a sphere can be approximated by a paraboloid of revolution
\begin{equation}
H  = h +\frac{r^2}{2 R} +O(r^4)\quad (r\ll R),
\label{H}
\end{equation}
where $r^2=x^2+y^2$.
The solution of Eq.(\ref{NS}) by imposing boundary conditions (\ref{BC_0}) and (\ref{BC_H}) yields
\begin{equation} \label{vtau}
\left(v_{\tau}\right)_i = \frac{\nabla_j p}{2\mu}\left[ z^2 \delta_{ij} - A_{ij}z - B_{ij} \right],
\end{equation}
\[ A_{ik}H+\left(b_{\rm eff}\right)_{ij} A_{jk}=H^2 \delta_{ik}, \quad B_{ik}=\left(b_{\rm eff}\right)_{ij}A_{jk},  \]
where $i$ and $j$  $=\{x,y\}$.

For convenience, we now diagonalize the effective slip length tensor $\textbf{b}_{\rm eff}$, by aligning $x$-axis with the `fast' axis of greatest forward slip, $b_{\rm eff}^{\parallel}$, which is always perpendicular to the `slow' axis of least forward slip, $b_{\rm eff}^{\perp}$~\cite{Bazant08}. By integrating Eq.(\ref{conteq}) with (\ref{BC_0})-(\ref{BC_H}), and by using (\ref{vtau}), we then derive a partial differential equation for pressure
\begin{equation}\label{velU}
 -\mu U=\frac{\partial}{\partial x}\left(H k_{\parallel}(H)\frac{\partial p}{\partial x}\right)+\frac{\partial}{\partial y}\left(H k_{\perp}(H)\frac{\partial p}{\partial y}\right),
\end{equation}
where $k_{\parallel,\perp}$ are the permeabilities of a \emph{flat} channel of thickness $H$ with one SH wall
\[ k_{\parallel,\perp}= \frac{H^2}{12}\left(\frac{H+4 b_{\rm eff}^{\parallel,\perp}(H)}{H+b_{\rm eff}^{\parallel,\perp}(H)}\right) = k \times k^{\ast}_{\parallel,\perp}\]
Here $k=H^2/12$ represents an \emph{isotropic} permeability of a flat hydrophilic channel with the same thickness, and $k^{\ast}_{\parallel,\perp}$ are corrections to this permeability due to SH slip.

The solution of Eq.(\ref{velU}) represents a very difficult problem since in general case, a broken symmetry of the problem caused by a texture anisotropy does not allow one standard simplifications, such as representing $p(x,y)$ as $p(H)$~\cite{vinogradova.oi:1995a,vinogradova:96}, etc.  Since $k_{\parallel ,\perp }$ are radially symmetric, it is more convenient to solve Eq.(\ref{velU}) by using polar coordinates
\begin{gather}
-\mu U=\frac{\partial }{\partial r}\left( H  \langle  k \rangle \frac{\partial p}{%
\partial r}\right) +\frac{H \langle k\rangle}{r}\frac{\partial p}{\partial r}+%
\frac{H \langle k\rangle}{r^{2}}\frac{\partial ^{2}p}{\partial \varphi ^{2}}
\label{a1} \\
+\cos 2\varphi \left[ \frac{\partial }{\partial r}\left(H \Delta k \frac{%
\partial p}{\partial r}\right) -\frac{H\Delta k }{r}\frac{\partial p}{%
\partial r}-\frac{H\Delta k }{r^{2}}\frac{\partial ^{2}p}{\partial
\varphi ^{2}}\right]   \notag \\
-\frac{\sin 2\varphi }{r}\left[ 2H \Delta k \frac{\partial ^{2}p}{%
\partial r\partial \varphi }+\left( \frac{d\left(H \Delta k \right) }{dr}%
-\frac{2H\Delta k }{r}\right) \frac{\partial p}{\partial \varphi }\right]
,  \notag
\end{gather}%
where
\begin{equation*}
 \langle k \rangle (r)=\frac{k _{\parallel}+k _{\perp}}{2},\quad \Delta k (r)=%
\frac{k_{\parallel}-k_{\perp}}{2},
\end{equation*}%
with boundary conditions
\begin{equation}
p(R,\varphi )=0,\quad \frac{\partial p}{\partial r}(0,\varphi )=0.
\label{bc}
\end{equation}%
The condition for a pressure takes the
form $p(r\rightarrow \infty )=0$, provided $R\gg \max \{b,L,h\}$.

In general case, the
solution of Eq.(\ref{a1}) is not radially symmetric, $\Delta k \neq 0,$ due to the terms
proportional to $\cos 2\varphi $ and $\sin 2\varphi$. However, it is
symmetric with respect to $x$ and $y$ axes, hence, it can be presented in
terms of cosine series:
\begin{equation} \label{Gener_sol_p}
p=p_{0}\left( r\right) +p_{1}\left( r\right) \cos 2\varphi +p_{2}\left(
r\right) \cos 4\varphi +...
\end{equation}%
This expansion can then be used to solve Eq. (\ref{a1}) numerically. Also an asymptotic solution can be constructed, provided  $\varepsilon = \Delta k \left( 0\right) / \langle k\rangle\left( 0\right) \ll 1$~\cite{epaps}.

The hydrodynamic resistance force is given by
\begin{equation}
F(h)= F_T f^{\ast} = \int\limits_{0}^{\infty }\int\limits_{0}^{2\pi }{p}r{\;d\varphi dr=}%
2\pi R\int\limits_{h}^{\infty }{p}_{0}{\;dH}.  \label{Force}
\end{equation}
We stress that terms with $\cos(2n\varphi)$ vanish after integration over $\varphi$ in the interval $[0, 2\pi]$, so that only \emph{isotropic} part of the pressure, $p_{0}$, contributes to the drag force.

In two limiting cases, namely small ($H\ll L$) and large ($H\gg L$)
distances between surfaces, the pressure distribution could be treated with high accuracy as radially symmetric~\cite{epaps}.
Once such assumption is made, the governing equation Eq.(\ref{a1}) for pressure could be rewritten in the form:%
\begin{equation}
-\mu U=\frac{d}{dr}\left(H \langle k \rangle\frac{d\tilde p}{dr}\right) +\frac{%
H \langle k \rangle }{r}\frac{d\tilde p}{dr}.  \label{der2P}
\end{equation}%
Here we use tilde to distinguish approximate $\tilde p$ from $p_0$, which is an isotropic part of a general solution, Eq.(\ref{Gener_sol_p}).
Eq.(\ref{der2P}) can be then integrated to give%
\begin{equation}
\frac{d\tilde p}{dr}=-\frac{\mu Ur}{2 H \langle k \rangle}=-\frac{12\mu Ur}{%
H^{3}\left(k_\parallel^{\ast }(H)+k_\perp^{\ast }(H)\right)},  \label{derivP}
\end{equation}%
so that we can represent $p(x,y)$ as a function of $H(x,y)$:
\begin{equation}
\frac{\tilde p(H)}{12 \mu UR}=\int\limits_{H}^{\infty }{\frac{ dH^{\prime }}{%
H^{\prime }{}^{3} \left(k_\parallel^{\ast }+k_\perp^{\ast }\right)}}.  \label{Pressure}
\end{equation}
The hydrodynamic resistance force can be then calculated as

\begin{equation}
F=F_{T} f^{\ast } = 2\pi R\int\limits_{h}^{\infty }\tilde{p}{\;dH}
\end{equation}%

In general case, the pressure and the force should be found numerically and their calculations require detailed knowledge about eigenvalues of effective slip tensor for some particular texture. However, in some situations asymptotic solutions could be obtained, and they are the same for all textures. Below we briefly discuss some of them.

\p For \emph{large} gap, $H \gg L$, the corrections to permeability are \cite{Bazant08}
\begin{equation}\label{isotr_per}
k_{\parallel }^{\ast }\simeq 1+\frac{3b_{\mathrm{eff}}^{\parallel }}{H}%
,\quad k_{\perp }^{\ast }\simeq 1+\frac{3b_{\mathrm{eff}}^{\perp }}{H}.
\end{equation}%
Even for strongly slipping ($b_2 \gg L$) surfaces, the effective slip lengths remain
independent of the gap thickness and are small compared to it~\cite{belyaev.av:2010b}. Consequently,
\begin{equation*}
\langle k\rangle=1+\frac{3(b_{\mathrm{eff}}^{\parallel
}+b_{\mathrm{eff}}^{\perp })}{H}+O\left( \varepsilon ^{2}\right),
\end{equation*}%
\begin{equation*}
\Delta k =\frac{3(b_{\mathrm{eff}}^{\parallel }-b_{\mathrm{eff}%
}^{\perp })}{H}\left[ 1+O\left( \varepsilon \right) \right] ,
\end{equation*}%
so that $\Delta k /\langle k \rangle \simeq b_{\mathrm{%
eff}}^{\parallel }/H\ll 1.$ The solution of Eq.(\ref{a1}) can then be constructed in terms of a series
of $\varepsilon$. In this case a non-axisymmetric part of the pressure field is small, and the axisymmetric
part may be approximated by $\tilde{p}$ with a very high accuracy, $p_{0}=%
\tilde{p}\left[ 1+O\left( \varepsilon ^{2}\right) \right]$~\cite{epaps}. The correction to SH slip is then
\begin{equation}
f^{\ast} \simeq 1-\frac{b_{\mathrm{eff}}^{\parallel }+b_{\mathrm{eff}%
}^{\perp }}{2 h} \label{asymp_large}
\end{equation}%
which coincides with the result, obtained earlier for corrugated (Wenzel) surfaces via Lorentz reciprocal
theorem~\cite{lecoq.n:2004}. Physically, Eq.(\ref{asymp_large}) means that the anisotropic SH plane at a distance $h$ from the sphere apex is equivalent to a no-slip
plane at a distance $h +(b_{\mathrm{eff}}^{\parallel }+b_{\mathrm{eff}}^{\perp })/2$. The shift of this equivalent no-slip plane from the real SH surface is thus equal to the average of the eigenvalues of the effective slip-length tensor.

For weakly slipping surfaces, $b_2 \ll L $, Eq.(\ref{asymp_large}) can be further simplified since in this situation the effective slip-length tensor is isotropic, and the effective slip coincides with the surface average~\cite{belyaev.av:2010a,kamrin.k:2010}. Therefore,

\begin{equation}
f^{\ast}\simeq 1-\frac{b_1 \phi_1 + b_2 \phi_2}{h} \label{asymp_large_iso}
\end{equation}%

\p For \emph{thin} gap, $H \ll L$, and weakly slipping surfaces, $b_2 \ll H $, the flow is also isotropic~\cite{feuillebois.f:2009}, and corrections to permeabilities are given by Eq.(\ref{isotr_per}). In what follows corrections for SH slip are given by Eq.(\ref{asymp_large_iso}). In other situations $f^{\ast}$ is sensitive to the pattern geometry and should be calculated individually for a particular texture of interest.

\section{Striped super-hydrophobic surfaces}

To  illustrate the predictions of the general theory, we now focus on a flat, periodic, striped SH
surface. This canonical texture is convenient to explore the basic physics of the system since the local
(scalar) slip lengths, $b_1$ and $b_2$, vary only in one direction, thus allowing us to highlight effects of anisotropy. The problem of flow past striped SH surfaces has previously been studied in a context of a reduction of pressure-driven forward
flow in thick~\cite{lauga.e:2003,belyaev.av:2010a,priezjev.nv:2005} and thin~\cite{feuillebois.f:2009,feuillebois.f:2010} channels,  and it is directly relevant for a passive
mixing~\cite{feuillebois.f:2010b,vinogradova.oi:2011}, and a generation of a tensorial electro-osmotic flow~\cite{bahga:2009,vinogradova.oi:2011}.

\subsection{Arbitrary gap, non-slipping solid regions}

The striped SH surface with $b_1=0$ currently seems to be only one where the eigenvalues of the slip length tensor have been calculated for an arbitrary thickness of the channel~\cite{vinogradova.oi:2011,schmieschek.s:2011}. For the general case the eigenvalues can be calculated semi-analytically, but accurate analytical results have been found in the thin and thick channel limits.

In the case of thick channels, $H \gg L$, the eigenvalues of the slip-length tensor read~\cite{belyaev.av:2010a}
\begin{equation}\label{beff_par_largeH}
  b_{\rm eff}^{\parallel} \simeq \frac{L}{\pi} \frac{\ln\left[\sec\left(\displaystyle\frac{\pi \phi_2}{2 }\right)\right]}{1+\displaystyle\frac{L}{\pi b_2}\ln\left[\sec\displaystyle\left(\frac{\pi \phi_2}{2 }\right)+\tan\displaystyle\left(\frac{\pi \phi_2}{2}\right)\right]},
\end{equation}
\begin{equation}\label{beff_ort_largeH}
  b_{\rm eff}^{\perp} \simeq \frac{L}{2 \pi} \frac{\ln\left[\sec\left(\displaystyle\frac{\pi \phi_2}{2 }\right)\right]}{1+\displaystyle\frac{L}{2 \pi b_2}\ln\left[\sec\displaystyle\left(\frac{\pi \phi_2}{2 }\right)+\tan\displaystyle\left(\frac{\pi \phi_2}{2}\right)\right]}.
\end{equation}
Flow in a large channel does not depend on $H$, and is controlled by the ratio of the local slip length $b$ to texture period $L$.
 When $b_2/L \ll 1$, Eqs.(\ref{beff_par_largeH}) and (\ref{beff_ort_largeH}) predict the area-averaged isotropic slip length, $b_{\rm eff}^{\perp, \parallel} \simeq \phi_2 b_2$. When $b/L \gg 1$,  expressions (\ref{beff_par_largeH}) and (\ref{beff_ort_largeH}) take form
\begin{equation}\label{beff_ort_largeH_id}
  b_{\rm eff}^{\perp} \simeq \frac{L}{2 \pi} \ln\left[\sec\left(\displaystyle\frac{\pi \phi_2}{2 }\right)\right],\quad b_{\rm eff}^{\parallel}\simeq2 b_{\rm eff}^{\perp},
\end{equation}
that coincides with results obtained for the perfect slip ($b_2 = \infty$) stripes~\cite{lauga.e:2003}.

 In the case of thin channels, $H \ll L$, striped surfaces have been shown to provide rigorous upper and lower Wiener bounds on the effective slip over all possible two-phase patterns~\cite{feuillebois.f:2009}
 \begin{equation}\label{b_stripes}
    b_{\rm eff}^{\parallel} = \frac{b H \phi_2}{H + b \phi_1},\quad b_{\rm eff}^{\perp} = \frac{b H \phi_2}{H + 4b \phi_1}
\end{equation}
At $b/H\gg 1$ these give truly tensorial anisotropic effective slip
\begin{equation}\label{beff_smallH_limit2}
  b_{\rm eff}^{\parallel} = H \frac{\phi_2}{\phi_1},\quad b_{\rm eff}^{\perp} = \frac{b_{\rm eff}^{\parallel}}{4},
\end{equation}
but at  $b/H\ll 1$ it
leads again to a surface average slip.

\begin{figure}
 \includegraphics[width=7.5 cm]{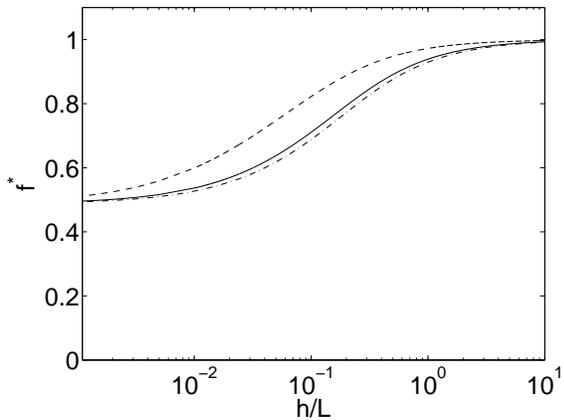}\\
   \caption{A correction to the resistance force for SH slip, $f^{\ast}$, as a function of $h/L$, calculated with $b_1=0$ and $b_2/L=10$ (dash-dots), $1$ (solid curve), and $0.1$ (dashed curve). The fraction of gas regions is $\phi_2=0.5$}
  \label{fig:force}
\end{figure}

Calculations of $b_{\rm eff}/H$, made for several $b_2/L$, were used to compute pressure, and then the correction for effective slip, $f^{\ast}$, as a function of the gap (Fig.~\ref{fig:force}). For this numerical example we have chosen $\phi_2=0.5$, which gives a maximum transverse flow in a thin channel situation~\cite{feuillebois.f:2010b}. All calculations were performed in the assumption of the radially symmetric pressure field, $p=\tilde{p}(H)$, which should give reasonably accurate results  even at the intermediate distances, i.e.  when the effective slip changes rapidly with $H$~\cite{epaps, vinogradova.oi:2011}. Fig.~\ref{fig:force} shows that at large distances all curves converge to $f^{\ast}=1$, i.e. the drag force is the same that it would be in case of a hydrophilic plane, $F=F_T$. This conclusion directly follows from Eq.~(\ref{asymp_large}) and is valid for any, however large, $b_2$. The correction for SH slip significantly decreases when $h$ becomes of the order of $L$ and smaller. It can be seen that the increase in $b_2$ leads to a smaller drag force at large and intermediate distances. However, at $h\ll L$ all curves tends to a constant, which does not depend on the value of the local slip length at the gas sectors.

\begin{figure}
 \includegraphics[width=7.5 cm]{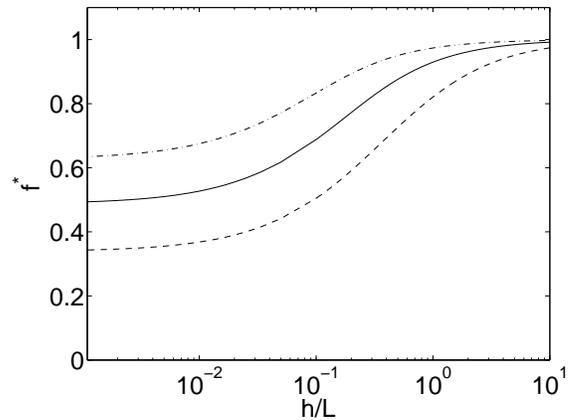}\\
  \caption{A correction for SH slip to the resistance force exerted on a sphere moving towards the anisotropic plane of stripes with no slip on solid-liquid regions ($b_1=0$).  $\phi_2=0.3$ (dash-dotted), $0.5$ (solid), $0.8$ (dash), $b_2/L=10$.}
  \label{fig:force1}
\end{figure}

Fig.~\ref{fig:force1} includes theoretical curves calculated for $b_2/L=10$ with different fractions of the surface gas phase. Results show that the correction for effective slip has a tendency to decrease with $\phi_2$, and that for each $\phi_2$ there is some minimum value of $f^{\ast }$ that all curves approach asymptotically. We are now on a position to quantify this important result. When $b_1=0$ and $h\ll \min \{b_2,L\}$ corrections to permeabilities take a form~\cite{feuillebois.f:2009}
\begin{equation}
k_{\parallel }^{\ast }=1+3\phi _{2},\quad k_{\perp }^{\ast }=\frac{4}{%
4-3\phi _{2}},
\end{equation}%
i.e. $k_{\parallel }^{\ast }/k_{\perp }^{\ast }$ is independent of
$H,$ and $\tilde{p}$ coincides with the exact solution of Eq. (\ref{a1}). \textrm{Therefore},
\begin{equation}
\frac{\tilde{p}\left( H\right) }{3\mu UR}=\frac{%
2(4-3\phi _{2})}{(8+9\phi _{2}-9\phi _{2}^{2})H^{2}}
\end{equation}%
and
\begin{equation}
f^{\ast }=\frac{2(4-3\phi _{2})}{8+9\phi _{2}-9\phi _{2}^{2}}  \label{b0}
\end{equation}%
 This result coincides with the expression obtained earlier for an interaction of a disk with a similar SH surface \cite{belyaev.av:2010b} and shows that in the limit of thin gap $f^{\ast }$ depends only on the surface fraction of the gas phase.

\begin{figure}
 \includegraphics[width=7 cm]{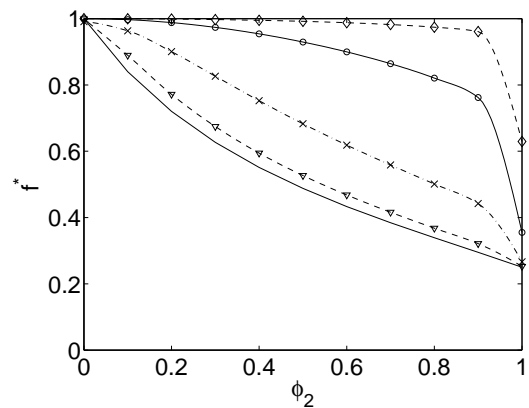}\\
  \caption{A correction for SH slip as a function of a fraction of the gas phase at $b_1=0$ and $b_2/L=10$. From top to bottom curves with symbols correspond to $h/L=0.01,$ 0.1, 1, and 10. Solid line without symbols shows calculation results obtained in the limit of a thin gap, $h \ll \min\{b_2,L\}$.}
  \label{fig:fstar_vs_phi2}
\end{figure}

Since $\phi_2$ is one of the key parameter determining
reduction of drag, in Fig.~\ref{fig:fstar_vs_phi2} we plot the correction for SH slip as a function of  $\phi_2$ at several $h/L$.
We see that when $\phi_2$ is very small, the correction for SH slip tends to its absolute
maximum, $f^{\ast }=1$. In the most interesting limit, $\phi_2 \to 1$, we can
reach the minimum possible value of correction for SH slip, $f^{\ast } \to 1/4$, provided $b_2/h$ is large.

\subsection{Thin gap, slipping solid regions}

In previous subsection we ignored the (small) slip length at the solid areas. The finite $b_1$ cannot modify our conclusions made for a thick or intermediate gap. It could, however, influence a thin gap situation. Fortunately,
 in the limit of a thin channel, the exact values of longitudinal and transverse permeabilities are known for any piecewise constant local slip length, $b_1$ and $b_2$. These values represent bounds that constrain the attainable effective permeabilities  and allows us to calculate effective slip lengths~\cite{feuillebois.f:2009}, and then $f^{\ast}$ for arbitrary local slip lengths.

The expression for a longitudinal permeability can be presented as~\cite{feuillebois.f:2009}
\begin{equation}\label{permeability_par}
k_\parallel = \phi_1 k_1 + \phi_2 k_2=k  \times k^{\ast}_{\parallel}
\end{equation}
with
\begin{equation}\label{k_par_star}
  k^{\ast}_{\parallel} = \phi_1 \left( \frac{H+4 b_1}{H + b_1} \right) + \phi_2 \left( \frac{H+4 b_2}{H + b_2} \right)
\end{equation}
Transverse stripes, in turn, satisfy the equation
\begin{equation}\label{permeability_perp}
 k_\perp  = \left( \frac{\phi_1}{k_1}+ \frac{\phi_2}{k_2} \right)^{-1}=k \times k^{\ast}_{\perp}
\end{equation}
where
\begin{equation}\label{k_perp_star}
  k^{\ast}_{\perp} = \left(\phi_1 \frac{H+b_1}{H+ 4 b_1} + \phi_2 \frac{H+b_2}{H+ 4 b_2} \right)^{-1}
\end{equation}
These expressions allow one to calculate effective slip lengths in eigendirections
\begin{equation}\label{parallel_b}
b_{\mathrm{eff}}^{\parallel }= \frac{H \phi_1 b_1 + H \phi_2 b_2 + b_1 b_2}{H+ \phi_2 b_1 + \phi_1 b_2}
\end{equation}%

\begin{equation}\label{transverse_b}
b_{\mathrm{eff}}^{\perp }= \frac{H \phi_1 b_1 + H \phi_2 b_2 + 4 b_1 b_2}{H+ 4 \phi_2 b_1 + 4 \phi_1 b_2}
\end{equation}%

In general case, the pressure and the force should be calculated numerically with the procedure described above. Some special cases, however, allow one exact analytical solutions.
In particular, when $H\ll \min \{b_2,L\}$ the corrections to permeabilities due to SH effective slip become
\begin{equation}\label{k_star1}
  k^{\ast}_{\parallel}(H) \simeq \frac{H (1+3\phi_2)+4 b_1}{H+b_1}
\end{equation}
\begin{equation}\label{k_star2}
  k^{\ast}_{\perp}(H)  \simeq \frac{4 (H + 4 b_1)}{4 (H+b_1) - 3 \phi_2 H}
\end{equation}

\begin{figure}
 \includegraphics[width=7.5 cm]{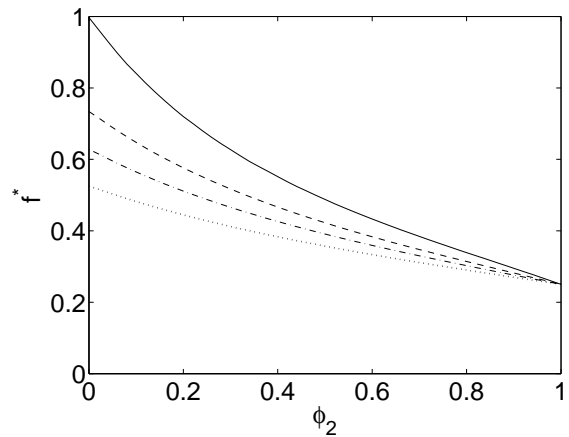}\\
  \caption{The correction for SH slip, $f^{\ast }$,  vs. surface fraction of the gas phase, $\phi _{2}$, in the limit of a thin gap and for $b_{2}/h\gg 1$. From top to bottom $b_{1}/h= 0$, $0.5$, $1$ and $2$.}
  \label{fig:fstar}
\end{figure}

The expressions for pressure and force in this situation are rigorously derived in Appendix~\ref{appendix2}, and an exact analytical expression for $f^{\ast}$ is given by Eq.(\ref{f_s}). Fig.~\ref{fig:fstar} includes theoretical curves, $f^{\ast}$ vs. $\phi _{2}$, calculated for several $b_{1}/h$. At $\phi _{2} \to 0,$ and $\phi _{2} \to 1$, our results are consistent with earlier predictions for homogeneous hydrophobic solids, with $b_1$ and $b_2$, described by Eq.(\ref%
{fb0})~\cite{vinogradova.oi:1995d}. In general, a finite hydrophobic slippage at the solid sectors can significantly reduce  $f^{\ast}$ at short distances.

\section{Final remarks}

Certain aspects of our work warrant further comments. We have presented data describing the squeeze-film
drainage of liquid in thin films between a hydrophilic sphere and a SH plane. The predicted drag force is smaller than expected for two hydrophilic solids and shows much more complex behavior as compared with expected in case of a uniform hydrophobic wall. For convenience, we now summarize our main results for striped SH surface in Table~\ref{table:1}. At first sight it is somewhat surprising that a system with one SH surface should exhibit such rich hydrodynamic properties. However, it becomes almost self-evident (from the presence of extra fundamental length scales, such as $L$, and two local slip lengths) that it should exhibit more rich behavior than a geometry with hydrophilic or even homogeneous hydrophobic surfaces. Do our results have any consequence for force experiments with real SH surfaces?

\begin{table}
  \centering
  \caption{Some useful asymptotic expressions for the correction factor, $f^{\ast}$, in case of a striped super-hydrophobic surface}
   \renewcommand{\baselinestretch}{2}\normalsize
\begin{tabular}{|c|c|}
\hline
  Case &  $f^{\ast}$ \\
  \hline
  $h\ll \min \{b_2,L\}$ & Eq.(\ref{f_s}) \\ \hline
  $b_1=0$, $h\ll \min \{b_2,L\}$ & $\displaystyle\frac{2(4-3\phi _{2})}{8+9\phi _{2}-9\phi _{2}^{2}}$ \\  \hline
  $L\gg h\gg b_{2}$ & $1- \displaystyle\frac{b_1 \phi_1+ b_2 \phi_2}{h}$ \\  \hline
  $h\gg L$ & $1-\displaystyle\frac{b_{\rm{eff}}^{\parallel}+b_{\rm{eff}}^{\perp }}{2h},$ \\& with Eqs.(\ref{beff_par_largeH}) and (\ref{beff_ort_largeH})\\
  \hline
\end{tabular}\label{table:1}
\end{table}

Hydrodynamic force can be measured using various techniques, such as AFM, SFA, and more. The geometry of configuration is equivalent to a sphere, of a large radius of curvature, approaching a flat surface. Hydrodynamic forces at small (compared to $R$) separations have been already reported for textured isotropic periodic~\cite{steinberger.a:2007} and random~\cite{wang.y:2010} SH surfaces. Also we are not aware of any measurements of the hydrodynamic force for the striped SH surface, measurements with isotropic textures lend some support to the picture of the significant reduction of a hydrodynamic force presented here. It would appear that reduction of a hydrodynamic drag that is observed is consistent with an earlier theoretical model, Eq.(\ref{fb0}), that gives $b_{\rm eff}$ of the order of 50-200 nm ~\cite{wang.y:2010,steinberger.a:2007}, which is only slightly larger than observed at smooth hydrophobic solids~\cite{vinogradova.oi:2009,vinogradova.oi:2003,honig.cdf:2007}. We suggest that further analysis of these measurements or similar measurements with other SH textures should employ one of the Eqs. given in Table~\ref{table:1} rather than Eq.(\ref{fb0}) since the latter makes assumptions of a constant, isotropic, homogeneous, and independent on distance slip which are not generally valid for a heterogeneous SH surface. While we maintain that Eq.(\ref{fb0}) is the appropriate formula for describing $f^{\ast}$ at very small ($\phi_2 \to 0$) and very large ($\phi_2 \to 1$) fractions of the gas phase we note that other formulas will be required in all different situations with the regime of validity given in Table~\ref{table:1}.

From our calculations it is evident that simple far-field Eq.(\ref{asymp_large}) gives an accurate estimate of $f^{\ast}$, and should certainly be used to deduce the average of eigenvalues of ${\bf b}_{\rm eff}$ at large separations. We note that similar evaluations could be
used for \emph{rough} or \emph{porous} surfaces since at large distances from the
wall the boundary condition at the rough interface or fluid-porous interface may be approximated by a slip model~\cite{lecoq.n:2004,vinogradova.oi:2006,kunert-harting-07,kunert.c:2010}. It should be possible to make such an analysis of recent data obtained with grooved surfaces~\cite{guriyanova.s:2010}, which we suspect will significantly alter the conclusion of these authors. Note that the average eigenvalues of an effective slip length tensor at large (compared to $L$) separations should coincide with the measured by velocimetry~\cite{tsai.p:2009} or other far-field methods~\cite{choi.ch:2006}. We remark and stress, however, that the near-field effective slip often measured in AFM/SFA should be much smaller than a far-field slip. This is exactly what has been observed in experiment~\cite{wang.y:2010,steinberger.a:2007}.

\iffalse%%%%%
\begin{figure}
 \includegraphics[width=6.5 cm]{asmolov_fig6.eps}\\
  \caption{Hydrodynamic force $F$ acting on a hydrophilic sphere of radius R approaching
 rough superhydrophobic stripes (solid curve) with $\phi_2=0.8$ and local slip length $b/L=100$. Dashed line is the Stokes
drag $F_{St} = 6\pi\mu UR$.}
  \label{fig:force1}
\end{figure}
\fi%%%%%

\section*{Acknowledgements}

We are grateful to F.~Feuillebois for helpful remarks on the manuscript and S.~Manuilovich for an advise concerning a numerical scheme. This research was supported by the RAS through its priority program `Assembly and Investigation of
Macromolecular Structures of New Generations' and by the DFG through its priority program `Micro- and nanofluidics'.

\vfil

\appendix

\section{Analytic solution for $\widetilde{p}$ and $f^{\ast }$ in the
thin-channel limit}\label{appendix2}

In this Appendix we derive analytical expressions for a dimensionless
pressure distribution $\widetilde{P}=\widetilde{p}h^{2}\left( 3\mu
UR\right) ^{-1}$ and for $f^{\ast }$ in a situation when $h\ll \min \{b_{2},L\}$ and the corrections to
effective permeabilities are given by Eqs. (\ref{k_star1}), (\ref{k_star2}).

We use the substitution into Eq. (\ref{Pressure}):%
\begin{equation*}
x=4-\frac{\eta ^{\prime }+4\beta }{\eta ^{\prime }+\beta },\quad \eta
^{\prime }=\frac{H^{\prime }}{h},\quad \beta =\frac{b_{1}}{h},
\end{equation*}%
to obtain%
\begin{widetext}
\begin{eqnarray}
\widetilde{P}(\eta ) &=&4\int\limits_{\eta }^{\infty }{\frac{d\eta
^{\prime }}{\eta ^{\prime }{}^{3}\left( k_{\parallel }^{\ast }+k_{\perp
}^{\ast }\right) }}  \notag \\
&=&4\int_{\frac{3\eta }{\eta +\beta }}^{3}\frac{\left( 3-x\right) \left(
4-\phi _{2}x\right) dx}{x^{3}\left[ \phi _{1}\phi _{2}\left( 4-x\right)
^{2}+4\left( \phi _{1}^{2}+\phi _{2}^{2}+1\right) \left( 4-x\right) +16\phi
_{1}\phi _{2}\right] }  \notag \\
&=&\frac{1}{4}\left. \left[ -\frac{3}{x^{2}}+\frac{\left( 4-3\phi
_{1}\right) }{2x}+\beta ^{2}c_{1}\ln \left\vert \frac{x^{2}}{\phi _{1}\phi
_{2}x^{2}-8x+32}\right\vert +\beta ^{2}c_{2}\ln \left\vert \frac{\phi
_{1}\phi _{2}x-4-d}{\phi _{1}\phi _{2}x-4+d}\right\vert \right] \right\vert
_{\frac{3\eta }{\eta +\beta }}^{3}  \notag \\
&=&\frac{3}{4}\left[ \frac{1}{3\eta ^{2}}+\frac{\phi _{1}}{2\beta \eta }%
+c_{1}\ln \left( \frac{\eta ^{2}+a_{1}\eta +a_{2}}{\eta ^{2}}\right)
+c_{2}\ln \left( \frac{\eta +a_{3}}{\eta +a_{4}}\right) \right] ,
\label{Pressure2}
\end{eqnarray}%
\begin{eqnarray*}
c_{1} &=&\frac{\phi _{1}\left( 3\phi _{1}-5\right) }{32\beta ^{2}},\quad
c_{2}=\frac{\phi _{1}\left( 3\phi _{1}-1\right) \left( 2\phi _{1}-3\right) }{%
8\beta ^{2}a_{6}}, \\
a_{1} &=&\frac{40\beta }{a_{5}},\quad a_{2}=\frac{32\beta ^{2}}{a_{5}},\quad
a_{3}=\frac{\beta \left( a_{6}-4\right) }{3\phi _{1}\phi _{2}+a_{6}-4},\  \\
a_{4} &=&-\frac{\beta \left( a_{6}+4\right) }{3\phi _{1}\phi _{2}-a_{6}-4}%
,\quad a_{5}=8+9\phi _{2}-9\phi _{2}^{2},\quad a_{6}=4\sqrt{1-2\phi
_{2}+2\phi _{2}^{2}}.
\end{eqnarray*}%
Then one can also integrate analytically Eq. (\ref{Force}) with $\widetilde{P}$
given by (\ref{Pressure2}) to get%
\begin{eqnarray}
f^{\ast } &=&\int_{1}^{\infty }\widetilde{P}d\eta   \notag \\
&=&\frac{3}{4}\left\{ -\frac{1}{3\eta }+\frac{\phi _{1}}{2\beta }\ln \eta
+c_{1}\left[ \left( \eta +\frac{a_{1}}{2}\right) \ln \left( \eta
^{2}+a_{1}\eta +a_{2}\right) +a_{7}\ln \left( \frac{2\eta +a_{1}+2a_{7}}{%
2\eta +a_{1}-2a_{7}}\right) -a_{1}-2\eta \ln \eta \right] \right.   \notag \\
&&+\left. \left. c_{2}\left[ \left( \eta +a_{3}\right) \ln \left( \eta
+a_{3}\right) -\left( \eta +a_{4}\right) \ln \left( \eta +a_{4}\right) %
\right] \right\} \right\vert _{1}^{\infty }  \notag \\
&=&\frac{3}{4}\left\{ \frac{1}{3}-c_{1}\left[ \left( 1+\frac{a_{1}}{2}%
\right) \ln \left( 1+a_{1}+a_{2}\right) -a_{1}\right] +\left[
c_{1}a_{7}+c_{2}\left( 1+a_{4}\right) \right] \ln \left( 1+a_{4}\right)
\right.   \notag \\
&&-\left. \left[ c_{1}a_{7}+c_{2}\left( 1+a_{3}\right) \right] \ln \left(
1+a_{3}\right) -c_{2}\left( a_{4}-a_{3}\right) \right\} ,  \label{f_s}
\end{eqnarray}%
\begin{equation*}
a_{7}=3\beta a_{6}/a_{5}.
\end{equation*}%
\end{widetext}

Eq. (\ref{f_s}) reduces to Eq. (\ref{fb0}) as $\phi _{2}\rightarrow 0$ and
to Eq. (\ref{b0}) as $\beta \rightarrow 0$.
\vfil

\bibliographystyle{rsc}
\bibliography{slipsphere1}

\providecommand*{\mcitethebibliography}{\thebibliography}
\csname @ifundefined\endcsname{endmcitethebibliography}
{\let\endmcitethebibliography\endthebibliography}{}
\begin{mcitethebibliography}{51}
\providecommand*{\natexlab}[1]{#1}
\providecommand*{\mciteSetBstSublistMode}[1]{}
\providecommand*{\mciteSetBstMaxWidthForm}[2]{}
\providecommand*{\mciteBstWouldAddEndPuncttrue}
  {\def\EndOfBibitem{\unskip.}}
\providecommand*{\mciteBstWouldAddEndPunctfalse}
  {\let\EndOfBibitem\relax}
\providecommand*{\mciteSetBstMidEndSepPunct}[3]{}
\providecommand*{\mciteSetBstSublistLabelBeginEnd}[3]{}
\providecommand*{\EndOfBibitem}{}
\mciteSetBstSublistMode{f}
\mciteSetBstMaxWidthForm{subitem}
{(\emph{\alph{mcitesubitemcount}})}
\mciteSetBstSublistLabelBeginEnd{\mcitemaxwidthsubitemform\space}
{\relax}{\relax}

\bibitem[Quere(2005)]{quere.d:2005}
D.~Quere, \emph{Rep. Prog. Phys.}, 2005, \textbf{68}, 2495--2532\relax
\mciteBstWouldAddEndPuncttrue
\mciteSetBstMidEndSepPunct{\mcitedefaultmidpunct}
{\mcitedefaultendpunct}{\mcitedefaultseppunct}\relax
\EndOfBibitem
\bibitem[Blossey(2003)]{blossey.r:2003}
R.~Blossey, \emph{Nature Materials}, 2003, \textbf{2}, 301--306\relax
\mciteBstWouldAddEndPuncttrue
\mciteSetBstMidEndSepPunct{\mcitedefaultmidpunct}
{\mcitedefaultendpunct}{\mcitedefaultseppunct}\relax
\EndOfBibitem
\bibitem[Richard and Quere(2000)]{richard.d:2000}
D.~Richard and D.~Quere, \emph{Europhys. Lett.}, 2000, \textbf{50},
  769--775\relax
\mciteBstWouldAddEndPuncttrue
\mciteSetBstMidEndSepPunct{\mcitedefaultmidpunct}
{\mcitedefaultendpunct}{\mcitedefaultseppunct}\relax
\EndOfBibitem
\bibitem[Tsai \emph{et~al.}(2010)Tsai, van~der Veen, van~de Raa, and
  Lohse]{tsai.p:2010}
P.~Tsai, R.~C.~A. van~der Veen, M.~van~de Raa and D.~Lohse, \emph{Langmuir},
  2010, \textbf{26}, 16090--16095\relax
\mciteBstWouldAddEndPuncttrue
\mciteSetBstMidEndSepPunct{\mcitedefaultmidpunct}
{\mcitedefaultendpunct}{\mcitedefaultseppunct}\relax
\EndOfBibitem
\bibitem[{Stone} \emph{et~al.}(2004){Stone}, {Stroock}, and
  {Ajdari}]{stone2004}
H.~A. {Stone}, A.~D. {Stroock} and A.~{Ajdari}, \emph{Annual Review of Fluid
  Mechanics}, 2004, \textbf{36}, 381--411\relax
\mciteBstWouldAddEndPuncttrue
\mciteSetBstMidEndSepPunct{\mcitedefaultmidpunct}
{\mcitedefaultendpunct}{\mcitedefaultseppunct}\relax
\EndOfBibitem
\bibitem[Squires and Quake(2005)]{squires2005}
T.~M. Squires and S.~R. Quake, \emph{Reviews of Modern Physics}, 2005,
  \textbf{77}, 977\relax
\mciteBstWouldAddEndPuncttrue
\mciteSetBstMidEndSepPunct{\mcitedefaultmidpunct}
{\mcitedefaultendpunct}{\mcitedefaultseppunct}\relax
\EndOfBibitem
\bibitem[Ybert \emph{et~al.}(2007)Ybert, Barentin, Cottin-Bizonne, Joseph, and
  Bocquet]{ybert.c:2007}
C.~Ybert, C.~Barentin, C.~Cottin-Bizonne, P.~Joseph and L.~Bocquet, \emph{Phys.
  Fluids}, 2007, \textbf{19}, 123601\relax
\mciteBstWouldAddEndPuncttrue
\mciteSetBstMidEndSepPunct{\mcitedefaultmidpunct}
{\mcitedefaultendpunct}{\mcitedefaultseppunct}\relax
\EndOfBibitem
\bibitem[Feuillebois \emph{et~al.}(2009)Feuillebois, Bazant, and
  Vinogradova]{feuillebois.f:2009}
F.~Feuillebois, M.~Z. Bazant and O.~I. Vinogradova, \emph{Phys. Rev. Lett.},
  2009, \textbf{102}, 026001\relax
\mciteBstWouldAddEndPuncttrue
\mciteSetBstMidEndSepPunct{\mcitedefaultmidpunct}
{\mcitedefaultendpunct}{\mcitedefaultseppunct}\relax
\EndOfBibitem
\bibitem[Vinogradova and Belyaev(2011)]{vinogradova.oi:2011}
O.~I. Vinogradova and A.~V. Belyaev, \emph{J. Phys.: Condens. Matter}, 2011,
  \textbf{23}, 184104\relax
\mciteBstWouldAddEndPuncttrue
\mciteSetBstMidEndSepPunct{\mcitedefaultmidpunct}
{\mcitedefaultendpunct}{\mcitedefaultseppunct}\relax
\EndOfBibitem
\bibitem[Vinogradova \emph{et~al.}(2009)Vinogradova, Koynov, Best, and
  Feuillebois]{vinogradova.oi:2009}
O.~I. Vinogradova, K.~Koynov, A.~Best and F.~Feuillebois, \emph{Phys. Rev.
  Lett.}, 2009, \textbf{102}, 118302\relax
\mciteBstWouldAddEndPuncttrue
\mciteSetBstMidEndSepPunct{\mcitedefaultmidpunct}
{\mcitedefaultendpunct}{\mcitedefaultseppunct}\relax
\EndOfBibitem
\bibitem[{Vinogradova} and {Yakubov}(2003)]{vinogradova.oi:2003}
O.~I. {Vinogradova} and G.~E. {Yakubov}, \emph{Langmuir}, 2003, \textbf{19},
  1227--1234\relax
\mciteBstWouldAddEndPuncttrue
\mciteSetBstMidEndSepPunct{\mcitedefaultmidpunct}
{\mcitedefaultendpunct}{\mcitedefaultseppunct}\relax
\EndOfBibitem
\bibitem[Cottin-Bizonne \emph{et~al.}(2005)Cottin-Bizonne, Cross, Steinberger,
  and Charlaix]{charlaix.e:2005}
C.~Cottin-Bizonne, B.~Cross, A.~Steinberger and E.~Charlaix, \emph{Phys. Rev.
  Lett.}, 2005, \textbf{94}, 056102\relax
\mciteBstWouldAddEndPuncttrue
\mciteSetBstMidEndSepPunct{\mcitedefaultmidpunct}
{\mcitedefaultendpunct}{\mcitedefaultseppunct}\relax
\EndOfBibitem
\bibitem[Squires(2008)]{Squires08}
T.~M. Squires, \emph{Phys.~Fluids}, 2008, \textbf{20}, 092105\relax
\mciteBstWouldAddEndPuncttrue
\mciteSetBstMidEndSepPunct{\mcitedefaultmidpunct}
{\mcitedefaultendpunct}{\mcitedefaultseppunct}\relax
\EndOfBibitem
\bibitem[Bahga \emph{et~al.}(2010)Bahga, Vinogradova, and Bazant]{bahga:2009}
S.~S. Bahga, O.~I. Vinogradova and M.~Z. Bazant, \emph{J.~Fluid Mech.}, 2010,
  \textbf{644}, 245--255\relax
\mciteBstWouldAddEndPuncttrue
\mciteSetBstMidEndSepPunct{\mcitedefaultmidpunct}
{\mcitedefaultendpunct}{\mcitedefaultseppunct}\relax
\EndOfBibitem
\bibitem[Belyaev and Vinogradova(2011)]{belyaev.av:2011a}
A.~V. Belyaev and O.~I. Vinogradova, \emph{Phys. Rev. Lett.}, 2011,
  submitted\relax
\mciteBstWouldAddEndPuncttrue
\mciteSetBstMidEndSepPunct{\mcitedefaultmidpunct}
{\mcitedefaultendpunct}{\mcitedefaultseppunct}\relax
\EndOfBibitem
\bibitem[Feuillebois \emph{et~al.}(2010)Feuillebois, Bazant, and
  Vinogradova]{feuillebois.f:2010b}
F.~Feuillebois, M.~Z. Bazant and O.~I. Vinogradova, \emph{Phys. Rev. E}, 2010,
  \textbf{82}, 055301(R)\relax
\mciteBstWouldAddEndPuncttrue
\mciteSetBstMidEndSepPunct{\mcitedefaultmidpunct}
{\mcitedefaultendpunct}{\mcitedefaultseppunct}\relax
\EndOfBibitem
\bibitem[Belyaev and Vinogradova(2010)]{belyaev.av:2010b}
A.~V. Belyaev and O.~I. Vinogradova, \emph{Soft Matter}, 2010, \textbf{6},
  4563--4570\relax
\mciteBstWouldAddEndPuncttrue
\mciteSetBstMidEndSepPunct{\mcitedefaultmidpunct}
{\mcitedefaultendpunct}{\mcitedefaultseppunct}\relax
\EndOfBibitem
\bibitem[not()]{note1}
Note that this famous formula never appeared in any of G. I. Taylor's
  publications as discussed in~\cite{horn.rg:2000}\relax
\mciteBstWouldAddEndPuncttrue
\mciteSetBstMidEndSepPunct{\mcitedefaultmidpunct}
{\mcitedefaultendpunct}{\mcitedefaultseppunct}\relax
\EndOfBibitem
\bibitem[Vinogradova(1995)]{vinogradova.oi:1995d}
O.~I. Vinogradova, \emph{J. Colloid Interface Sci.}, 1995, \textbf{169},
  306--319\relax
\mciteBstWouldAddEndPuncttrue
\mciteSetBstMidEndSepPunct{\mcitedefaultmidpunct}
{\mcitedefaultendpunct}{\mcitedefaultseppunct}\relax
\EndOfBibitem
\bibitem[Vinogradova(1995)]{vinogradova.oi:1995a}
O.~I. Vinogradova, \emph{Langmuir}, 1995, \textbf{11}, 2213\relax
\mciteBstWouldAddEndPuncttrue
\mciteSetBstMidEndSepPunct{\mcitedefaultmidpunct}
{\mcitedefaultendpunct}{\mcitedefaultseppunct}\relax
\EndOfBibitem
\bibitem[Chan and Horn(1985)]{chan.dyc:1985}
D.~Y.~C. Chan and R.~G. Horn, \emph{J. Chem. Phys.}, 1985, \textbf{83},
  5311\relax
\mciteBstWouldAddEndPuncttrue
\mciteSetBstMidEndSepPunct{\mcitedefaultmidpunct}
{\mcitedefaultendpunct}{\mcitedefaultseppunct}\relax
\EndOfBibitem
\bibitem[not()]{note2}
Direct approaches to flow profiling, or velocimetry, take advantage of various
  optics to monitor tracer particles. Their accuracy is normally much lower
  than that of force methods due to relatively low optical resolution, system
  noise due to polydispersity of tracers, and difficulties in decoupling flow
  from diffusion.\relax
\mciteBstWouldAddEndPunctfalse
\mciteSetBstMidEndSepPunct{\mcitedefaultmidpunct}
{}{\mcitedefaultseppunct}\relax
\EndOfBibitem
\bibitem[Honig and Ducker(2007)]{honig.cdf:2007}
C.~D.~F. Honig and W.~A. Ducker, \emph{Phys. Rev. Lett.}, 2007, \textbf{98},
  028305\relax
\mciteBstWouldAddEndPuncttrue
\mciteSetBstMidEndSepPunct{\mcitedefaultmidpunct}
{\mcitedefaultendpunct}{\mcitedefaultseppunct}\relax
\EndOfBibitem
\bibitem[Wang and Bhushan(2010)]{wang.y:2010}
Y.~Wang and B.~Bhushan, \emph{Soft Matter}, 2010, \textbf{6}, 29--66\relax
\mciteBstWouldAddEndPuncttrue
\mciteSetBstMidEndSepPunct{\mcitedefaultmidpunct}
{\mcitedefaultendpunct}{\mcitedefaultseppunct}\relax
\EndOfBibitem
\bibitem[Andrienko \emph{et~al.}(2004)Andrienko, Patricio, and
  Vinogradova]{andrienko.d:2004}
D.~Andrienko, P.~Patricio and O.~I. Vinogradova, \emph{J.~Chem Phys.}, 2004,
  \textbf{121}, 4414--4423\relax
\mciteBstWouldAddEndPuncttrue
\mciteSetBstMidEndSepPunct{\mcitedefaultmidpunct}
{\mcitedefaultendpunct}{\mcitedefaultseppunct}\relax
\EndOfBibitem
\bibitem[Yakubov \emph{et~al.}(2000)Yakubov, Butt, and Vinogradova]{yakubov:00}
G.~E. Yakubov, H.~J. Butt and O.~I. Vinogradova, \emph{J. Phys. Chem. B}, 2000,
  \textbf{104}, 3407 -- 3410\relax
\mciteBstWouldAddEndPuncttrue
\mciteSetBstMidEndSepPunct{\mcitedefaultmidpunct}
{\mcitedefaultendpunct}{\mcitedefaultseppunct}\relax
\EndOfBibitem
\bibitem[Tyrrell and Attard(2001)]{tyrrell.jwg:2001}
J.~W.~G. Tyrrell and P.~Attard, \emph{Phys. Rev. Lett.}, 2001, \textbf{87},
  176104\relax
\mciteBstWouldAddEndPuncttrue
\mciteSetBstMidEndSepPunct{\mcitedefaultmidpunct}
{\mcitedefaultendpunct}{\mcitedefaultseppunct}\relax
\EndOfBibitem
\bibitem[Parker \emph{et~al.}(1994)Parker, Claesson, and
  Attard]{parker.jl:1994}
J.~L. Parker, P.~M. Claesson and P.~Attard, \emph{J. Phys. Chem.}, 1994,
  \textbf{98}, 8468\relax
\mciteBstWouldAddEndPuncttrue
\mciteSetBstMidEndSepPunct{\mcitedefaultmidpunct}
{\mcitedefaultendpunct}{\mcitedefaultseppunct}\relax
\EndOfBibitem
\bibitem[Bazant and Vinogradova(2008)]{Bazant08}
M.~Z. Bazant and O.~I. Vinogradova, \emph{J.~Fluid Mech.}, 2008, \textbf{613},
  125--134\relax
\mciteBstWouldAddEndPuncttrue
\mciteSetBstMidEndSepPunct{\mcitedefaultmidpunct}
{\mcitedefaultendpunct}{\mcitedefaultseppunct}\relax
\EndOfBibitem
\bibitem[Kamrin \emph{et~al.}(2010)Kamrin, Bazant, and Stone]{kamrin.k:2010}
K.~Kamrin, M.~Bazant and H.~A. Stone, \emph{J. Fluid Mech.}, 2010,
  \textbf{658}, 409--437\relax
\mciteBstWouldAddEndPuncttrue
\mciteSetBstMidEndSepPunct{\mcitedefaultmidpunct}
{\mcitedefaultendpunct}{\mcitedefaultseppunct}\relax
\EndOfBibitem
\bibitem[Hyv\"{a}luoma and Harting(2008)]{harting.j:2008}
J.~Hyv\"{a}luoma and J.~Harting, \emph{Phys. Rev. Lett.}, 2008, \textbf{100},
  246001\relax
\mciteBstWouldAddEndPuncttrue
\mciteSetBstMidEndSepPunct{\mcitedefaultmidpunct}
{\mcitedefaultendpunct}{\mcitedefaultseppunct}\relax
\EndOfBibitem
\bibitem[Davis and Lauga(2009)]{lauga2009}
A.~M.~J. Davis and E.~Lauga, \emph{Phys. Fluids}, 2009, \textbf{21},
  011701\relax
\mciteBstWouldAddEndPuncttrue
\mciteSetBstMidEndSepPunct{\mcitedefaultmidpunct}
{\mcitedefaultendpunct}{\mcitedefaultseppunct}\relax
\EndOfBibitem
\bibitem[Sbragaglia and Prosperetti(2007)]{sbragaglia.m:2007}
M.~Sbragaglia and A.~Prosperetti, \emph{Phys. Fluids}, 2007, \textbf{19},
  043603\relax
\mciteBstWouldAddEndPuncttrue
\mciteSetBstMidEndSepPunct{\mcitedefaultmidpunct}
{\mcitedefaultendpunct}{\mcitedefaultseppunct}\relax
\EndOfBibitem
\bibitem[Pirat \emph{et~al.}(2008)Pirat, Sbragaglia, Peters, Borkent,
  Lammertink, Wessling, and Lohse]{pirat.c:2008}
C.~Pirat, M.~Sbragaglia, A.~M. Peters, B.~M. Borkent, R.~G.~H. Lammertink,
  M.~Wessling and D.~Lohse, \emph{Europhys. Lett.}, 2008, \textbf{81},
  6602\relax
\mciteBstWouldAddEndPuncttrue
\mciteSetBstMidEndSepPunct{\mcitedefaultmidpunct}
{\mcitedefaultendpunct}{\mcitedefaultseppunct}\relax
\EndOfBibitem
\bibitem[Reyssat \emph{et~al.}(2008)Reyssat, Yeomans, and
  Quere]{reyssat.m:2008}
M.~Reyssat, J.~M. Yeomans and D.~Quere, \emph{Europhys. Lett.}, 2008,
  \textbf{81}, 26006\relax
\mciteBstWouldAddEndPuncttrue
\mciteSetBstMidEndSepPunct{\mcitedefaultmidpunct}
{\mcitedefaultendpunct}{\mcitedefaultseppunct}\relax
\EndOfBibitem
\bibitem[Vinogradova(1996)]{vinogradova:96}
O.~I. Vinogradova, \emph{Langmuir}, 1996, \textbf{12}, 5963 -- 5968\relax
\mciteBstWouldAddEndPuncttrue
\mciteSetBstMidEndSepPunct{\mcitedefaultmidpunct}
{\mcitedefaultendpunct}{\mcitedefaultseppunct}\relax
\EndOfBibitem
\bibitem[epa()]{epaps}
See EPAPS document No. [number will be inserted by publisher] for details of
  analysis and a numerical scheme. For more information on EPAPS, see
  http://www.aip.org/pubservs/epaps.html\relax
\mciteBstWouldAddEndPuncttrue
\mciteSetBstMidEndSepPunct{\mcitedefaultmidpunct}
{\mcitedefaultendpunct}{\mcitedefaultseppunct}\relax
\EndOfBibitem
\bibitem[Lecoq \emph{et~al.}(2004)Lecoq, Anthore, Cichocki, Szymczak, and
  Feuillebois]{lecoq.n:2004}
N.~Lecoq, R.~Anthore, B.~Cichocki, P.~Szymczak and F.~Feuillebois, \emph{J.
  Fluid Mech.}, 2004, \textbf{513}, 247--264\relax
\mciteBstWouldAddEndPuncttrue
\mciteSetBstMidEndSepPunct{\mcitedefaultmidpunct}
{\mcitedefaultendpunct}{\mcitedefaultseppunct}\relax
\EndOfBibitem
\bibitem[Belyaev and Vinogradova(2010)]{belyaev.av:2010a}
A.~V. Belyaev and O.~I. Vinogradova, \emph{J.~Fluid Mech.}, 2010, \textbf{652},
  489--499\relax
\mciteBstWouldAddEndPuncttrue
\mciteSetBstMidEndSepPunct{\mcitedefaultmidpunct}
{\mcitedefaultendpunct}{\mcitedefaultseppunct}\relax
\EndOfBibitem
\bibitem[Lauga and Stone(2003)]{lauga.e:2003}
E.~Lauga and H.~A. Stone, \emph{J.~Fluid Mech.}, 2003, \textbf{489},
  55--77\relax
\mciteBstWouldAddEndPuncttrue
\mciteSetBstMidEndSepPunct{\mcitedefaultmidpunct}
{\mcitedefaultendpunct}{\mcitedefaultseppunct}\relax
\EndOfBibitem
\bibitem[Priezjev \emph{et~al.}(2005)Priezjev, Darhuber, and
  Troian]{priezjev.nv:2005}
N.~V. Priezjev, A.~A. Darhuber and S.~M. Troian, \emph{Phys. Rev. E}, 2005,
  \textbf{71}, 041608\relax
\mciteBstWouldAddEndPuncttrue
\mciteSetBstMidEndSepPunct{\mcitedefaultmidpunct}
{\mcitedefaultendpunct}{\mcitedefaultseppunct}\relax
\EndOfBibitem
\bibitem[Feuillebois \emph{et~al.}(2010)Feuillebois, Bazant, and
  Vinogradova]{feuillebois.f:2010}
F.~Feuillebois, M.~Z. Bazant and O.~I. Vinogradova, \emph{Phys. Rev. Lett.},
  2010, \textbf{104}, 159902\relax
\mciteBstWouldAddEndPuncttrue
\mciteSetBstMidEndSepPunct{\mcitedefaultmidpunct}
{\mcitedefaultendpunct}{\mcitedefaultseppunct}\relax
\EndOfBibitem
\bibitem[Schmieschek \emph{et~al.}(2011)Schmieschek, Belyaev, Harting, and
  Vinogradova]{schmieschek.s:2011}
S.~Schmieschek, A.~V. Belyaev, J.~Harting and O.~I. Vinogradova, \emph{Phys.
  Fluids}, 2011,  in preparation\relax
\mciteBstWouldAddEndPuncttrue
\mciteSetBstMidEndSepPunct{\mcitedefaultmidpunct}
{\mcitedefaultendpunct}{\mcitedefaultseppunct}\relax
\EndOfBibitem
\bibitem[Steinberger \emph{et~al.}(2007)Steinberger, Cottin-Bizonne, Kleimann,
  and Charlaix]{steinberger.a:2007}
A.~Steinberger, C.~Cottin-Bizonne, P.~Kleimann and E.~Charlaix, \emph{Nature
  Materials}, 2007, \textbf{6}, 665--668\relax
\mciteBstWouldAddEndPuncttrue
\mciteSetBstMidEndSepPunct{\mcitedefaultmidpunct}
{\mcitedefaultendpunct}{\mcitedefaultseppunct}\relax
\EndOfBibitem
\bibitem[Vinogradova and Yakubov(2006)]{vinogradova.oi:2006}
O.~I. Vinogradova and G.~E. Yakubov, \emph{Phys. Rev. E}, 2006, \textbf{73},
  045302(R)\relax
\mciteBstWouldAddEndPuncttrue
\mciteSetBstMidEndSepPunct{\mcitedefaultmidpunct}
{\mcitedefaultendpunct}{\mcitedefaultseppunct}\relax
\EndOfBibitem
\bibitem[Kunert and Harting(2007)]{kunert-harting-07}
C.~Kunert and J.~Harting, \emph{Phys. Rev. Lett}, 2007, \textbf{99},
  176001\relax
\mciteBstWouldAddEndPuncttrue
\mciteSetBstMidEndSepPunct{\mcitedefaultmidpunct}
{\mcitedefaultendpunct}{\mcitedefaultseppunct}\relax
\EndOfBibitem
\bibitem[Kunert \emph{et~al.}(2010)Kunert, Harting, and
  Vinogradova]{kunert.c:2010}
C.~Kunert, J.~Harting and O.~I. Vinogradova, \emph{Phys. Rev. Lett}, 2010,
  \textbf{105}, 016001\relax
\mciteBstWouldAddEndPuncttrue
\mciteSetBstMidEndSepPunct{\mcitedefaultmidpunct}
{\mcitedefaultendpunct}{\mcitedefaultseppunct}\relax
\EndOfBibitem
\bibitem[Guriyanova \emph{et~al.}(2010)Guriyanova, Semin, Rodrigues, Butt, and
  Bonaccurso]{guriyanova.s:2010}
S.~Guriyanova, B.~Semin, T.~S. Rodrigues, H.~J. Butt and E.~Bonaccurso,
  \emph{Microfluid Nanofluid}, 2010, \textbf{8}, 653--663\relax
\mciteBstWouldAddEndPuncttrue
\mciteSetBstMidEndSepPunct{\mcitedefaultmidpunct}
{\mcitedefaultendpunct}{\mcitedefaultseppunct}\relax
\EndOfBibitem
\bibitem[Tsai \emph{et~al.}(2009)Tsai, Peters, Pirat, Wessling, Lammerting, and
  Lohse]{tsai.p:2009}
P.~Tsai, A.~M. Peters, C.~Pirat, M.~Wessling, R.~G.~H. Lammerting and D.~Lohse,
  \emph{Phys. Fluids}, 2009, \textbf{21}, 112002\relax
\mciteBstWouldAddEndPuncttrue
\mciteSetBstMidEndSepPunct{\mcitedefaultmidpunct}
{\mcitedefaultendpunct}{\mcitedefaultseppunct}\relax
\EndOfBibitem
\bibitem[Choi \emph{et~al.}(2006)Choi, Ulmanella, Kim, Ho, and
  Kim]{choi.ch:2006}
C.~H. Choi, U.~Ulmanella, J.~Kim, C.~M. Ho and C.~J. Kim, \emph{Phys. Fluids},
  2006, \textbf{18}, 087105\relax
\mciteBstWouldAddEndPuncttrue
\mciteSetBstMidEndSepPunct{\mcitedefaultmidpunct}
{\mcitedefaultendpunct}{\mcitedefaultseppunct}\relax
\EndOfBibitem
\bibitem[Horn \emph{et~al.}(2000)Horn, Vinogradova, Mackay, and
  Phan-Thien]{horn.rg:2000}
R.~G. Horn, O.~I. Vinogradova, M.~E. Mackay and N.~Phan-Thien, \emph{J. Chem.
  Phys.}, 2000, \textbf{112}, 6424 -- 6433\relax
\mciteBstWouldAddEndPuncttrue
\mciteSetBstMidEndSepPunct{\mcitedefaultmidpunct}
{\mcitedefaultendpunct}{\mcitedefaultseppunct}\relax
\EndOfBibitem
\end{mcitethebibliography}


\providecommand*{\mcitethebibliography}{\thebibliography}
\csname @ifundefined\endcsname{endmcitethebibliography}
{\let\endmcitethebibliography\endthebibliography}{}
\begin{mcitethebibliography}{3}
\providecommand*{\natexlab}[1]{#1}
\providecommand*{\mciteSetBstSublistMode}[1]{}
\providecommand*{\mciteSetBstMaxWidthForm}[2]{}
\providecommand*{\mciteBstWouldAddEndPuncttrue}
  {\def\EndOfBibitem{\unskip.}}
\providecommand*{\mciteBstWouldAddEndPunctfalse}
  {\let\EndOfBibitem\relax}
\providecommand*{\mciteSetBstMidEndSepPunct}[3]{}
\providecommand*{\mciteSetBstSublistLabelBeginEnd}[3]{}
\providecommand*{\EndOfBibitem}{}
\mciteSetBstSublistMode{f}
\mciteSetBstMaxWidthForm{subitem}
{(\emph{\alph{mcitesubitemcount}})}
\mciteSetBstSublistLabelBeginEnd{\mcitemaxwidthsubitemform\space}
{\relax}{\relax}

\bibitem[Godunov(1961)]{godunov1961nsb}
S.~K. Godunov, \emph{Uspekhi Matematicheskikh Nauk}, 1961, \textbf{16},
  171--174\relax
\mciteBstWouldAddEndPuncttrue
\mciteSetBstMidEndSepPunct{\mcitedefaultmidpunct}
{\mcitedefaultendpunct}{\mcitedefaultseppunct}\relax
\EndOfBibitem
\bibitem[Conte(1966)]{conte1966nsl}
S.~D. Conte, \emph{SIAM Review}, 1966, \textbf{8}, 309\relax
\mciteBstWouldAddEndPuncttrue
\mciteSetBstMidEndSepPunct{\mcitedefaultmidpunct}
{\mcitedefaultendpunct}{\mcitedefaultseppunct}\relax
\EndOfBibitem
\bibitem[Feuillebois \emph{et~al.}(2009)Feuillebois, Bazant, and
  Vinogradova]{feuillebois.f:2009}
F.~Feuillebois, M.~Z. Bazant and O.~I. Vinogradova, \emph{Phys. Rev. Lett.},
  2009, \textbf{102}, 026001\relax
\mciteBstWouldAddEndPuncttrue
\mciteSetBstMidEndSepPunct{\mcitedefaultmidpunct}
{\mcitedefaultendpunct}{\mcitedefaultseppunct}\relax
\EndOfBibitem
\end{mcitethebibliography}

\end{document}

\title{\textbf{Supplementary information.\\ \emph{Drag force on a sphere moving towards
an anisotropic super-hydrophobic plane}}}

\author{Evgeny S. Asmolov}

\author{Aleksey V. Belyaev}

\author{Olga I. Vinogradova}

\maketitle

\section{Numerical and asymptotic solutions of equation for pressure}

Eq.(11) of our main paper has the following form:
\begin{gather}
-\mu U=\frac{\partial }{\partial r}\left( H  \langle  k \rangle \frac{\partial p}{%
\partial r}\right) +\frac{H \langle k\rangle}{r}\frac{\partial p}{\partial r}+%
\frac{H \langle k\rangle}{r^{2}}\frac{\partial ^{2}p}{\partial \varphi ^{2}}
\label{a1} \\
+\cos 2\varphi \left[ \frac{\partial }{\partial r}\left(H \Delta k \frac{%
\partial p}{\partial r}\right) -\frac{H\Delta k }{r}\frac{\partial p}{%
\partial r}-\frac{H\Delta k }{r^{2}}\frac{\partial ^{2}p}{\partial
\varphi ^{2}}\right]   \notag \\
-\frac{\sin 2\varphi }{r}\left[ 2H \Delta k \frac{\partial ^{2}p}{%
\partial r\partial \varphi }+\left( \frac{d\left(H \Delta k \right) }{dr}%
-\frac{2H\Delta k }{r}\right) \frac{\partial p}{\partial \varphi }\right]
,  \notag
\end{gather}%
and its solution can be presented in
terms of cosine series:
\begin{equation} \label{Gener_sol_p}
p=p_{0}\left( r\right) +p_{1}\left( r\right) \cos 2\varphi +p_{2}\left(
r\right) \cos 4\varphi +...
\end{equation}%

Here we derive equations for functions $p_{n}\left( r\right) $
in series Eq.(\ref{Gener_sol_p}) and describe their numerical solution. When $\Delta k$ is small, we construct an asymptotic solution.

To solve Eq.(\ref{a1}) we substitute (%
\ref{Gener_sol_p}) into (\ref{a1}) and collect terms proportional to $\cos
2n\varphi $. To eliminate singularities, we then use logarithmic
substitution for variable $r$ and introduce the following dimensionless
quantities:
\begin{eqnarray*}
\xi  &=&\ln \left( \frac{r}{\sqrt{hR}}\right) ,\quad \eta =\frac{H}{h}=1+%
\frac{\exp \left( 2\xi \right) }{2}, \\
P_{n} &=&\frac{p_{n}h^{2}}{3\mu UR}, \\
G\left( \eta \right)  &=&-\frac{\eta ^{3} \langle k \rangle}{\langle k%
\rangle\left( 0\right) },\quad D\left( \eta \right) =-\frac{\eta ^{3}\
\Delta k}{\Delta k\left( 0\right) }
\end{eqnarray*}%
This leads to a system of ordinary differential
equations
\begin{equation}
\mathrm{L}_{0}P_{0}+\varepsilon \mathrm{L}_{0}^{+}P_{1}=\frac{4 k }{\langle k%
\rangle\left( 0\right) }\exp \left( 2\xi \right) ,  \label{a5}
\end{equation}%
\begin{equation}
\mathrm{L}_{n}P_{n}+\varepsilon \mathrm{L}_{n}^{+}P_{n+1}+\varepsilon
\mathrm{L}_{n}^{-}P_{n-1}=0\quad \text{for}\quad n>0,  \label{a6}
\end{equation}%
which is expressed in a more compact form by using differential operators
\begin{equation*}
\mathrm{L}_{n}=\frac{d}{d\xi }\left( G\frac{d}{d\xi }\right) -4n^{2}G,
\end{equation*}%
\begin{eqnarray*}
\mathrm{L}_{n}^{+} &=&\frac{d}{d\xi }\left( \frac{D}{2}\frac{d}{d\xi }%
\right) +\left( 2n+1\right) D\frac{d}{d\xi } \\
&&+\left( n+1\right) \frac{dD}{d\xi }+2n\left( n+1\right) D,
\end{eqnarray*}%
\begin{eqnarray*}
\mathrm{L}_{n}^{-} &=&\frac{d}{d\xi }\left( \frac{D}{2}\frac{d}{d\xi }%
\right) -\left( 2n-1\right) D\frac{d}{d\xi } \\
&&-\left( n-1\right) \frac{dD}{d\xi }+2n\left( n-1\right) D,
\end{eqnarray*}%

\begin{figure}[tbp]
\includegraphics[width=7 cm]{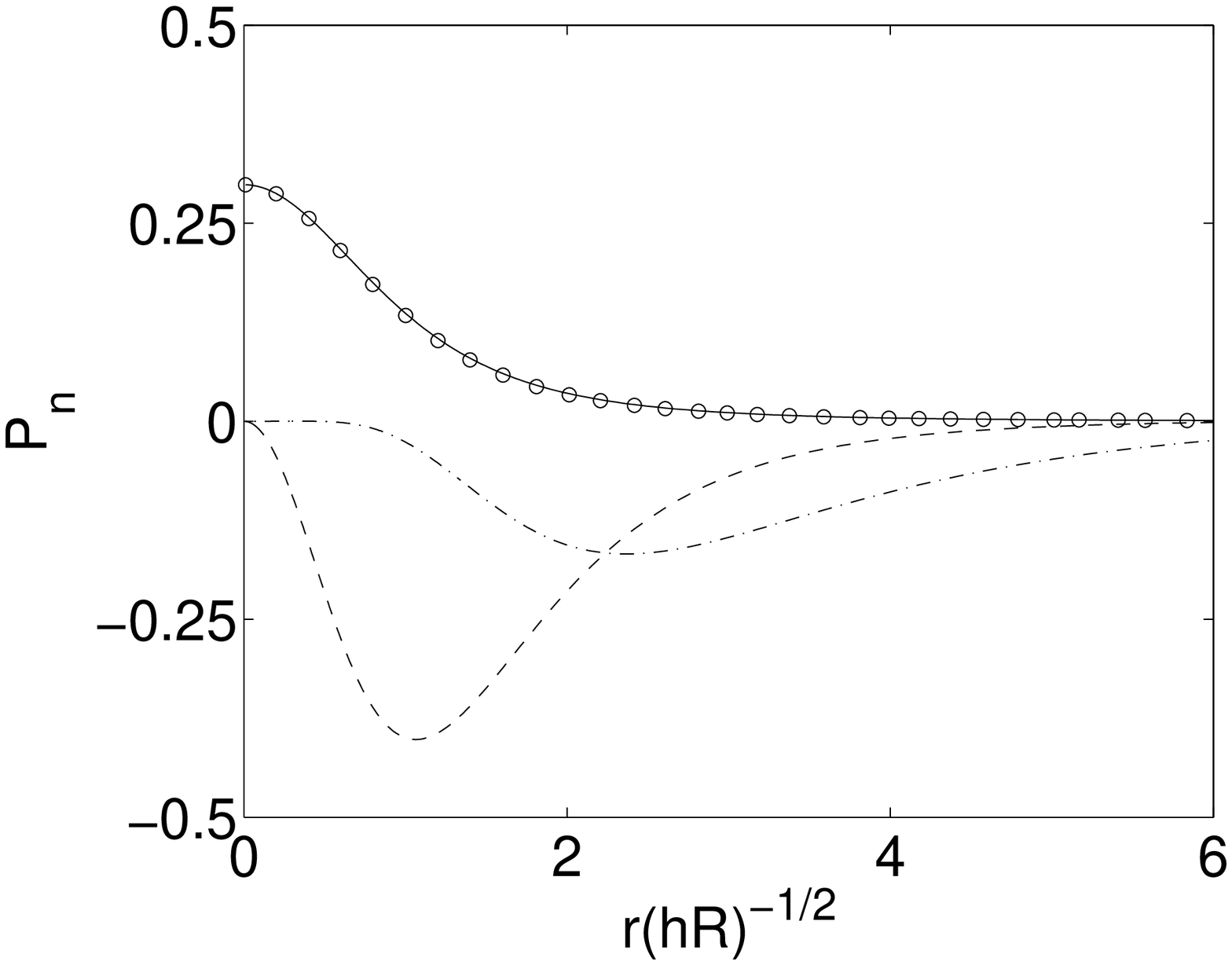}\newline
\caption{ Functions $P_{n}\left( r\right) =p_{n}h^{2}/3\protect\mu UR$ in
series (\protect\ref{Gener_sol_p}) for $h\ll \min \{ b_2,L \},$ $%
b_{1}/h=0.5$, $\protect\phi _{2}=0.8$. Solid line corresponds to an analytical
solution for $\tilde{P}=\tilde{p} h^{2}/3\protect\mu UR$, circles - to
a numerical evaluation of $P_0$, dashed line - to $10^3P_1$ and dash-dotted
line - to $10^5P_2$.}
\label{pn}
\end{figure}

To solve a finite-difference version of ODE system (\ref{a5}), (\ref{a6}) we
resolve the truncated system with respect to $d^{2}P_{n}/d\xi ^{2},$ $%
n=1,...,\ N,$ by using the Gauss routine, and then integrate numerically the obtained system by
using the fourth-order Runge-Kutta method. This system has a boundary condition
$P_{n}\rightarrow 0$ both at small and large $r,$ or at $\xi \rightarrow -\infty
$ and $\xi \rightarrow +\infty$. Therefore, the dimensionless gap and
the permeabilities take a form%
\begin{equation*}
\eta \rightarrow 1,\quad dG/d\xi =dD/d\xi \rightarrow 0\quad \text{as}\quad
\xi \rightarrow -\infty ,
\end{equation*}%
\begin{equation*}
\eta \simeq \frac{\exp \left( 2\xi \right) }{2},\quad G,\ D\simeq \eta
^{3}\simeq \exp \left( 6\xi \right) \quad \text{as}\quad \xi \rightarrow
+\infty .
\end{equation*}%
Asymptotic linearly independent solutions of (\ref{a5}), (\ref%
{a6}) can be then found in the form: $P_{ni}=a_{ni}^{0}\exp \left( \lambda
_{i}^{0}\xi \right) ,$ $i=1,...,\ 2N$ as $\xi \rightarrow -\infty $ and $%
P_{ni}=a_{ni}^{\infty }\exp \left( \lambda _{i}^{\infty }\xi \right) $ as $%
\xi \rightarrow +\infty .$ The eigenvalues, $\lambda _{i}^{0}$ and $\lambda
_{i}^{\infty },$ and eigenvectors, $\mathbf{a}_{i}^{0}$ and $\mathbf{a}%
_{i}^{\infty },$ were found using IMSL routine DEVCRG. The system admits
both exponentially growing and decaying solutions $P_{ni}$ whereas the
boundary conditions require the decaying one. Thus, we choose solutions with
$\lambda _{i}^{0}>0$ and $\lambda _{i}^{\infty }<0$. To resolve all linearly
independent solutions properly at $\left\vert \xi \right\vert \gg 1$ the
orthonormalization method~\cite{godunov1961nsb,conte1966nsl} is applied.
The boundaries of the integration domain are set at $r_{\min }=0.01$ and $%
r_{\max }=20,$ the number of taken into account harmonics  is $N=4.$ The
calculations show that the expansion converges fast with $N,$ since the
ratio $\left\vert P_{n+1}/P_{n}\right\vert $ is usually small for all $n$, which is illustrated in Fig.~\ref{pn}.

The solution of Eq. (\ref{a1}) is simple if ratio $k_{\parallel }/k_{\perp }$ (and hence $\Delta k/ \langle k \rangle$) is
constant, i.e., when the permeabilities depend on $r$ similarly, $%
k_{\parallel }^{\ast },\ k_{\perp }^{\ast }\sim g\left( r\right) .$ This is the case in a thin channel with $%
b_{1}=0$, when all the permeabilities are proportional to $H^{2}$~\cite%
{feuillebois.f:2009}. Then the solution includes only axisymmetric term $%
p_{0}$. It can be verified directly that solution
\begin{equation}
\frac{d\tilde p}{dr}=-\frac{\mu Ur}{2 H \langle k \rangle}=-\frac{12\mu Ur}{%
H^{3}\left(k_\parallel^{\ast }(H)+k_\perp^{\ast }(H)\right)},  \label{derivP}
\end{equation}%
\begin{equation}
\frac{\tilde p(H)}{12 \mu UR}=\int\limits_{H}^{\infty }{\frac{ dH^{\prime }}{%
H^{\prime }{}^{3} \left(k_\parallel^{\ast }+k_\perp^{\ast }\right)}}  \label{Pressure}
\end{equation}
satisfies Eq. (\ref{a1}) since the boundary conditions%
are homogeneous, $\partial \widetilde{p}/\partial \varphi =0$, and
terms
\begin{equation*}
\frac{\partial }{\partial r}\left(H \Delta k \frac{\partial \widetilde{p}%
}{\partial r}\right) =\frac{\Delta k}{2 \langle k \rangle}\mu
U,\quad \frac{H\Delta k }{r}\frac{\partial \widetilde{p}}{\partial r}%
=\frac{\Delta k}{2 \langle k \rangle}\mu U,
\end{equation*}%
cancel out. The analytical expressions for $\widetilde{P}=\widetilde{p}%
h^{2}(3\mu UR)^{-1}$ and corresponding resistance force in the case when
$h\ll \min \{ b_{2},L \} $ are obtained in the Appendix section of the main paper.

The first three harmonics in series (\ref{Gener_sol_p}), plotted as
functions of $r/\sqrt{hR}$, are presented in Fig.~\ref{pn}. It shows that
the axisymmetric part of pressure distribution is very close to the
approximate solution $\widetilde{P}$ while the non-axisymmetric part is
several orders less.

The permeability difference is typically small, $|\Delta k|\ll |\langle k\rangle|$, for example,
for large distances between surfaces, $H\gg L.$ The asymptotic solution of
Eqs. (\ref{a5}), (\ref{a6}) can be found in this important case as power
series expansion $P_{n}=\sum\nolimits_{j=0}\varepsilon ^{j}P_{n}^{(j)}$ with
respect to small parameter $\varepsilon $. We do not construct the full
asymptotic solution, but show that the isotropic part $P_{0}$, which only
contributes to the drag force, differs from the approximate solution $%
\widetilde{P}$ given by Eq. (\ref{Pressure}) by the value $O\left(
\varepsilon ^{2}\right) .$ Substituting the expansions into the system (\ref%
{a5}), (\ref{a6}) and collecting the terms proportional to $\varepsilon ^{j}$
one obtains equations governing the first three terms of expansion:
\begin{equation}
\mathrm{L}_{0}P_{0}^{(0)}=\frac{4 k}{\langle k\rangle\left( 0\right) }\exp
\left( 2\xi \right) ,  \label{a7}
\end{equation}%
\begin{equation*}
P_{n}^{(0)}=0\quad \text{as}\quad n>0,
\end{equation*}%
\begin{equation*}
\mathrm{L}_{1}P_{1}^{(1)}=-\mathrm{L}_{1}^{-}P_{0}^{(0)},
\end{equation*}%

\begin{equation*}
P_{n}^{(1)}=0\quad \text{as}\quad n\neq 1,
\end{equation*}%
\begin{equation*}
\mathrm{L}_{0}P_{0}^{(2)}=-\mathrm{L}_{0}^{+}P_{1}^{(1)},
\end{equation*}%
\begin{equation*}
\mathrm{L}_{2}P_{2}^{(2)}=-\mathrm{L}_{2}^{-}P_{1}^{(1)},
\end{equation*}%
\begin{equation*}
P_{1}^{(2)}=0,\ P_{n}^{(2)}=0\quad \text{as}\quad n>2.
\end{equation*}%
The equation (\ref{a7}) is the dimensionless form of Eq. (\ref{derivP}). For
further terms, one can deduce that
\begin{equation*}
P_{n}=\varepsilon ^{n}P_{n}^{(n)}+\varepsilon
^{n+2}P_{n}^{(n+2)}+\varepsilon ^{n+4}P_{n}^{(n+4)}+...
\end{equation*}%
Thus the first-order correction to the pressure field is $\varepsilon
P_{1}^{(1)},$ while the correction to the isotropic part $\varepsilon
^{2}P_{0}^{(2)}$ is much smaller.

\iffalse%%%%%
In the most practical situations the permeability ratio $\left\vert \Delta
k/\langle k\rangle\right\vert $ is small and/or weakly dependent
on $r$. As a result of both factors, numerical solution coincides well with
approximate solution $\widetilde{P}_{0}$ (see Fig.~\ref{pn}). \fi%%%%%

\bibliographystyle{rsc}
\bibliography{slipsphere1}